\documentclass[fleqn,useAMS,usenatbib]{mnras}

\usepackage{psfrag,color,epsfig}

\usepackage{graphicx}	
\usepackage{amsmath}	
\usepackage{amssymb}	
\usepackage{multicol}	
\usepackage{bm}		
\usepackage{pdflscape}	
\usepackage{multirow}
\usepackage{tabularx}  

\usepackage{todonotes}

\def\HI{\ion{H}{I}~}

\def\xHI{x_{\rm \ion{H}{I}}}
\def\xb{\bar{x}_{\rm \ion{H}{I}}}

\def\kk{{\bm{k}}}

\def\x{{\bm{x}}}

\newcommand{\be}{\begin{equation}}
\newcommand{\ba}{\begin{eqnarray}}
\newcommand{\ee}{\end{equation}}
\newcommand{\ea}{\end{eqnarray}}

\def\gtsima{$\; \buildrel > \over \sim \;$}
\def\ltsima{$\; \buildrel < \over \sim \;$}
\def\simgt{\lower.5ex\hbox{\gtsima}}
\def\simlt{\lower.5ex\hbox{\ltsima}}
\def\simpr{\lower.5ex\hbox{\prosima}}

\usepackage[T1]{fontenc}
\usepackage{ae,aecompl}

\usepackage{newtxtext,newtxmath}

\usepackage{tikz,xcolor,hyperref}
\definecolor{lime}{HTML}{A6CE39}
\DeclareRobustCommand{\orcidicon}{%
	\begin{tikzpicture}
	\draw[lime, fill=lime] (0,0) 
	circle [radius=0.16] 
	node[white] {{\fontfamily{qag}\selectfont \tiny ID}};
	\draw[white, fill=white] (-0.0625,0.095) 
	circle [radius=0.007];
	\end{tikzpicture}
	\hspace{-2mm}
}
\foreach \x in {A, ..., Z}{%
	\expandafter\xdef\csname orcid\x\endcsname{\noexpand\href{https://orcid.org/\csname orcidauthor\x\endcsname}{\noexpand\orcidicon}}
}

\title[LC Bispectrum \& detectability]{The Epoch of Reionization 21-cm Bispectrum: The impact of light-cone effects and detectability}

\author[R. Mondal et al.]{Rajesh Mondal\orcidA,$^1$\thanks{E-mail: \href{mailto:rajesh@astro.su.se}{rajesh@astro.su.se}} Garrelt Mellema\orcidB,$^1$  Abinash Kumar Shaw,$^2$ Mohd Kamran,$^3$ Suman Majumdar$^{3,4}$  
\\
$^{1}$The Oskar Klein Centre, Department of Astronomy, Stockholm University, AlbaNova, SE-10691 Stockholm, Sweden\\
$^{2}$Department of Physics, Indian Institute of Technology Kharagpur, Kharagpur 721 302, India\\
$^{3}$Department of Astronomy, Astrophysics and Space Engineering, Indian Institute of Technology Indore, Simrol, Indore 453552, India\\
$^{4}$Department of Physics, Blackett Laboratory, Imperial College, London SW7 2AZ, UK}

\date{Accepted 2021 October 5. Received 2021 October 5; in original form 2021 July 6.}
\pubyear{2021}

\begin{document}
\label{firstpage}
\pagerange{\pageref{firstpage}--\pageref{lastpage}}
\maketitle


\begin{abstract}
We study the spherically averaged bispectrum of the 21-cm signal from the Epoch of Reionization (EoR). This metric provides a quantitative measurement of the level of non-Gaussianity of the signal which is expected to be high. We focus on the impact of the light-cone effect on the bispectrum and its detectability with the future SKA-Low telescope. Our investigation is based on a single reionization light-cone model and an ensemble of 50 realisations of the 21-cm signal to estimate the cosmic variance errors. We calculate the bispectrum with a new, optimised direct estimation method, \textsc{DviSukta} which calculates the bispectrum for all possible unique triangles. We find that the light-cone effect becomes important on scales $k_1 \la 0.1\,{\rm Mpc}^{-1}$ where for most triangle shapes the cosmic variance errors dominate. Only for the squeezed limit triangles, the impact of the light-cone effect exceeds the cosmic variance. Combining the effects of system noise and cosmic variance we find that $\sim 3\sigma$ detection of the bispectrum is possible for all unique triangle shapes around a scale of $k_1 \sim 0.2\,{\rm Mpc}^{-1}$, and cosmic variance errors dominate above and noise errors below this length scale. Only the squeezed limit triangles are able to achieve a more than $5\sigma$ significance over a wide range of scales, $k_1\la 0.8\,{\rm Mpc}^{-1}$. Our results suggest that among all the possible triangle combinations for the bispectrum, the squeezed limit one will be the most measurable and hence useful.

\end{abstract}



\section{Introduction}
\label{sec:intro}
After the Big Bang the Universe expanded and gradually cooled until during the Epoch of Recombination the electrons and protons combined into neutral hydrogen and radiation decoupled from matter. After this period the Universe remained dark until the first luminous structures formed, a phase commonly known as the Cosmic Dawn\,(CD). These first luminous sources emitted copious amounts of ionizing radiation, gradually reionizing the \HI in the Inter-Galactic Medium\,(IGM). This period is therefore known as the Epoch of Reionization\,(EoR). Due to the paucity of observations, our understanding of the EoR remains limited \citep[see e.g. the recent introductory review in][]{2019ConPh..60..145W}.


The redshifted 21-cm signal, arising due to the hyperfine transition of the electron-proton system from parallel to anti-parallel spin in the ground state of \ion{H}{I}, is a powerful probe of astrophysical and cosmological information during the EoR \citep[see e.g.][] {furlanetto06,pritchard12}. Therefore a number of low frequency radio interferometers such as 
GMRT\footnote{\url{http://www.gmrt.ncra.tifr.res.in}} \citep{swarup91},  
LOFAR\footnote{\url{http://www.lofar.org/}} \citep{haarlem13}, 
MWA\footnote{\url{http://www.mwatelescope.org/}} \citep{tingay13}, 
PAPER \citep{parsons14} have been dedicating substantial efforts and amounts of observing time to the detection of the fluctuations in the 21-cm signal from the EoR. The relative weakness of the signal with respect to both system noise (see e.g. \citealt{morales05,mcquinn06}) and strong foreground emission, $\sim 4 - 5$ order of magnitude larger than the expected signal (see e.g. \citealt{ali08,ghosh12}), poses great challenges to these efforts. In spite of this, these first generation radio interferometers have successfully put significant upper limits on the expected EoR 21-cm signal \citep{barry19, li19, kolopanis19, Mertens2020, trott20}. The next generation of telescopes, HERA\footnote{\url{https://reionization.org/}} \citep{deboer17} and 
SKA\footnote{\url{http://www.skatelescope.org/}} \citep{koopmans15}, once completed, should through their much higher sensitivity be able to detect the fluctuations in the 21-cm signal from EoR much more easily.

The Spherically Averaged Power Spectrum (SAPS) provides an estimate of the fluctuations (variance) at different length scales. It quantifies the amplitudes of the fluctuations in the signal at different wavenumbers $k$. It fully describes the statistical properties of a field which consists of Gaussian random fluctuations. However, the underlying non-linear matter density field and especially the formation of extended ionized regions which contain no signal, introduce a high level of non-Gaussianity in the EoR $21$-cm signal \citep{bharadwaj05a, mellema06, mondal15} which implies that the power spectrum does not fully characterise the signal \citep{mondal16, mondal17}. One point statistics such as the skewness and kurtosis (see e.g. \citealt{Harker2009,watkinson14,Watkinson2015,shimabukuro15,kubota16}) do quantify the non-Gaussianity but do not describe its scale dependence. For this we require higher-order statistics such as the bispectrum (see e.g. \citealt{Peebles_book,Fry82,Fry99,Hivon95,Matarrese97,Scoccimarro97}).

The bispectrum is the Fourier transform of the three-point correlation function and therefore is a function of three distances, which can also be characterised by one scale factor and the chosen shape of a triangle. In the context of the EoR 21-cm signal, the non-Gaussianity was first studied using the Spherically Averaged Bispectrum (SABS) by \citet{bharadwaj05} using an analytical model consisting of spherical ionized regions. These authors first reported that the bispectrum can attain both positive and negative values. \citet{watkinson17} also confirmed the bispectrum sign. Using a suite of semi-numerical simulations, \citet{majumdar18} estimated the EoR 21-cm SABS for some specific triangles (e.g. equilateral, isosceles). They showed that the competition between matter density and neutral fraction fields decides the sign of the bispectrum. The bispectrum is negative when the non-Gaussianity is arising due to fluctuations in the neutral fraction whereas it is positive when the non-Gaussianity is caused by the matter density fluctuations. \citet{hutter19} independently observed similar kinds of features in their study of the 21-cm bispectrum. \citet{shimabukuro16} presented another independent study of the EoR 21-cm bispectrum. However, their estimator is unable to capture the sign of the bispectrum.

These earlier work are all based on studies of some specific shapes of triangles. The first comprehensive study of the EoR 21-cm bispectrum of all possible triangles was performed by \citet{majumdar20}. For this they used the prescription of all possible unique triangles in Fourier space developed by \citet{bharadwaj20}. A subsequent study of \citet{kamran21} presented the similar study for the 21-cm bispectrum from the CD. These studies showed that among all possible unique triangles, the limiting squeezed bispectrum typically has the largest magnitude.

To interpret the cosmological 21-cm observations we need to characterize their statistical properties, such as the SAPS and SABS. However, the statistical properties of a line transition signal such as the 21-cm signal change along the line-of-sight (LoS) direction since different frequencies originate from different look back times. This is known as the Light-Cone (LC) effect \citep{barkana06}. It has a particularly significant impact on the measured statistics when the mean of the signal changes rapidly with redshift. The impact of the LC effect on the EoR 21-cm SAPS has been considered in several studies (see e.g. \citealt{datta12, datta14, laplante14, mondal18, Greig2018}). These works have shown that the LC effect significantly affects the amplitude of the large scale 3D Fourier modes (i.e. small $k$) but mainly averages out at small scales (i.e. large $k$).

In this paper we will consider the impact of the LC effect on the bispectrum, an aspect which has not been considered before. We will work in the same framework as \citet{majumdar20} did. In analogy to the SAPS we expect the largest impact at large scales which is also where cosmic variance (CV) affects the measurements. Hence we include a study of the CV. Furthermore, we consider the detectability of the SABS for all possible unique triangles in future SKA-low observations by including a numerical noise calculation. Both CV and system noise have previously only been considered through approximations and/or for a limited set of triangle shapes (see e.g. \citealt{yoshiura2015, watkinson19, ma21, watkinson21}).

The structure of the paper is as follows. In Section~\ref{sec:bispec} we describe the theoretical formalism and the algorithm that we use to estimate the SABS from a simulated 21-cm signal for all possible unique triangles. Section~\ref{sec:sim} briefly describes our method to generate simulated 21-cm signals. In section \ref{sec:result} we discuss our main findings regarding the impact of LC effect. Following this, Section~\ref{sec:sensitivity} explores how well the future SKA-Low will be able to measure the EoR 21-cm SABS for all possible triangles, considering both cosmic variance and system noise. Finally, in Section~\ref{sec:conclusion} we summarise our findings.

Throughout this paper, we have used the Planck+WP best fit values of the cosmological parameters, viz. $h = 0.6704$, $\Omega_{\mathrm{m}0} = 0.3183$, $\Omega_{\Lambda0} = 0.6817$, $\Omega_{\mathrm{b}0} h^2 = 0.022032$,
$\sigma_8 = 0.8347$ and $n_s = 0.9619$ \citep{planck14}.


\section{The Spherically Averaged Bispectrum}
\label{sec:bispec}
The bispectrum $B(\kk_1, \kk_2, \kk_3)$ is defined through
\begin{equation}
\langle \Delta(\kk_1) \Delta(\kk_2) \Delta(\kk_3) \rangle= \delta_{\kk_1 + \kk_2 + \kk_3,0}~V\,B(\kk_1, \kk_2, \kk_3)\, , 
\label{eq:bispec1}
\end{equation}
where $\Delta(\kk)$ is the 3D Fourier transform of the fluctuations, $V$ is the comoving volume under consideration and $\langle...\rangle$ denotes the ensemble average. The Konecker delta $\delta_{\kk_1 + \kk_2 + \kk_3,0}$ is 1 if $\kk_1 + \kk_2 + \kk_3=0$ i.e. when the three $\kk$-vectors form a closed triangle, and 0 if not.

\subsection{All unique triangle configurations}
\label{sec:unique}
\begin{figure}
\centering
\includegraphics[width=0.48\textwidth, angle=0]{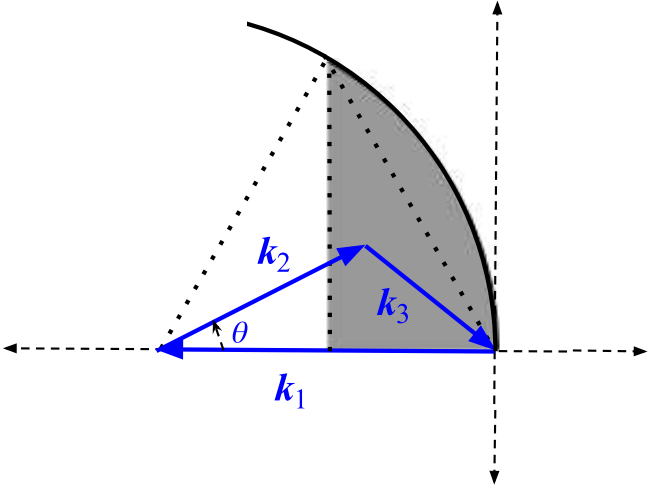}
\caption{All unique triangle configurations for a $\kk_1$-vector on a 2D plane. The grey shaded region corresponds to all unique triangle configurations which satisfy Eq.~(\ref{eq:triangle4})}
\label{fig:triangle}
\end{figure}

To find all possible unique closed triangle configurations in Fourier space we use the parameterization proposed by \citet{bharadwaj20}. In this formalism the size and shape of a triangle are quantified by identifying $\kk_1$ as the largest and $\kk_2$ the second largest side of the triangle (Fig.~\ref{fig:triangle}). This implies 
\begin{equation}
k_1 \ge k_2 \ge k_3\,,
\label{eq:triangle}
\end{equation}
where $k$ represents the amplitude of the $\kk$-vector. The amplitude $k_1$ quantifies the size of the triangle and the shape is quantified using the following two parameters
\begin{equation}
\begin{array}{l}
n=\frac{k_2}{k_1} \,, \\
\cos{\theta}=-\frac{\kk_1 \cdot \kk_2}{k_1k_2} \,, 
\end{array}
\label{eq:triangle2}
\end{equation}
where $\theta$ is the angle between $-\kk_1$ and $\kk_2$ vectors. Using Eq.~(\ref{eq:triangle}) in Eq.~(\ref{eq:triangle2}), we can derive that 
\begin{equation}
0.5 \le n \le 1\,,
\label{eq:triangle3}
\end{equation}
and
\begin{equation}
n\,\cos{\theta} \ge 0.5\,.
\label{eq:triangle4}
\end{equation}
The shaded region in Fig.~\ref{fig:triangle} shows all unique triangle configurations for a $\kk_1$-vector on a 2D plane which satisfy Eq.~(\ref{eq:triangle4}).

\subsection{The direct estimator of the bispectrum}
We use Eq.~(\ref{eq:bispec1}) to define the binned SABS estimator 
\begin{equation}
\hat{B}(k_1,n,\cos{\theta}) \equiv \hat{B}(k_1,k_2,k_3) = \frac{1}{N_{\rm t}V} \sum_i \Delta(\kk_1) \Delta(\kk_2) \Delta(\kk_3)\,,
\label{eq:est}
\end{equation}
where the sum $\sum_i$ is over $N_{\rm t}$ number of closed triangles within the $i$-th bin. Note that the bins here are three dimensional (3D) voxels of volume $[\Delta k_1\,\Delta k_2\,\Delta k_3]$ which we map to the $(k_1,n,\cos{\theta})$ space using Eqs.~(\ref{eq:triangle}) and (\ref{eq:triangle2}). The ensemble average of the estimator gives the bin-averaged SABS $\langle \hat{B} (k_1, n, \cos{\theta}) \rangle = \bar{B} (k_1, n, \cos{\theta})$.

In a conventional direct estimation method, we can directly use Eq.~(\ref{eq:est}) on gridded data in Fourier space to estimate the SABS. If the data consists of $N_{\rm G}^3$ grid points, we can restrict the $\kk_1$ search to half of those i.e. to $N_{\rm G}^3/2$ grid points in Fourier space, as the modes $\kk_{\rm i}$ and $-\kk_{\rm i}$ give the same estimates of the SABS. Therefore, about $N_{\rm G}^6/2$ operations are required to evaluate the condition $\kk_3=-(\kk_1+\kk_2)$ and estimate the SABS, as one needs to search $\kk_2$ over all $N_{\rm G}^3$ grid points for $N_{\rm G}^3/2$ number of $\kk_1$. Hence, the computing time increases very steeply with $N_{\rm G}$.

To decrease this large number of operations, one can use a fast estimator based on the Fast Fourier transform (FFT) (see e.g. \citealt{watkinson17, shaw21}). However, the associated increase in speed does come with some disadvantages. The most important of these is that the data needs to be in a form appropriate for FFT, so the data set has to be periodic and cannot contain any gaps, conditions which are not typically fulfilled for real data (see e.g. \citealt{trott19}). Furthermore, it cannot estimate the polyspectrum of order $p$ on scales $k > 2\pi/p\Delta L$, where $\Delta L$ is the resolution. A second complication is that it is not straightforward to convert the derived polyspectrum which is a function of $k_{\rm i}$ to another representation such as our parametrization for all unique triangles which uses $[k_1, n, \cos{\theta}]$ space. A last drawback arises from the fact that the FT based estimator estimates a $k$-bin averaged polyspectrum, and cannot preserve the information of the orientation of individual $\kk$-vectors with respect to the LoS direction within a bin. Therefore, it is difficult to quantify the polyspectrum's anisotropy in terms of the multipole moments \citep{bharadwaj20} using this method. However, this particular aspect we will not consider in the current work, but plan to study anisotropies in the future.

To incorporate all unique triangle configurations, \citet{majumdar18} proposed a restricted implementation of the direct estimation method. Their method uses Eq.~(\ref{eq:triangle2}) and calculates the SABS at specified values of $n$ and $\cos{\theta}$. This eliminates two nested $f\!or$-loops from the direct triangle search algorithm, and reduces the number of operations to $N_{\rm n} N_{\cos{\theta}} N_{\rm G}^4/2$, when the bispectrum is calculated for $N_{\rm n}$ and $N_{\cos{\theta}}$ numbers of $n$ and $\cos{\theta}$ values, respectively. However, $N_{\rm n}$ and $N_{\cos{\theta}}$ are fixed and do not depend on the number of grid points in the input data. This makes the algorithm very restrictive in nature as it does not allow any kind of bin width around $n$ and $\cos{\theta}$. Their method can also suffer from a sampling bias, when for a given $\kk_1$ the number of available grid points in the allowed region (the shaded region in Fig.~\ref{fig:triangle}) turns out to be less than $N_{\rm n} N_{\cos{\theta}}$ which means that these values of the SABS will be over-sampled. This is not a severe problem for large values of $\kk_1$, where the number of Fourier modes is very large. However, this is a significant disadvantage at small $\kk_1$, where radio interferometers actually have most of their sensitivity\footnote{We discuss the sensitivity of radio interferometers to the SABS in Section~\ref{sec:sensitivity}.}.

\begin{figure}
\centering
\includegraphics[width=0.48\textwidth, angle=0]{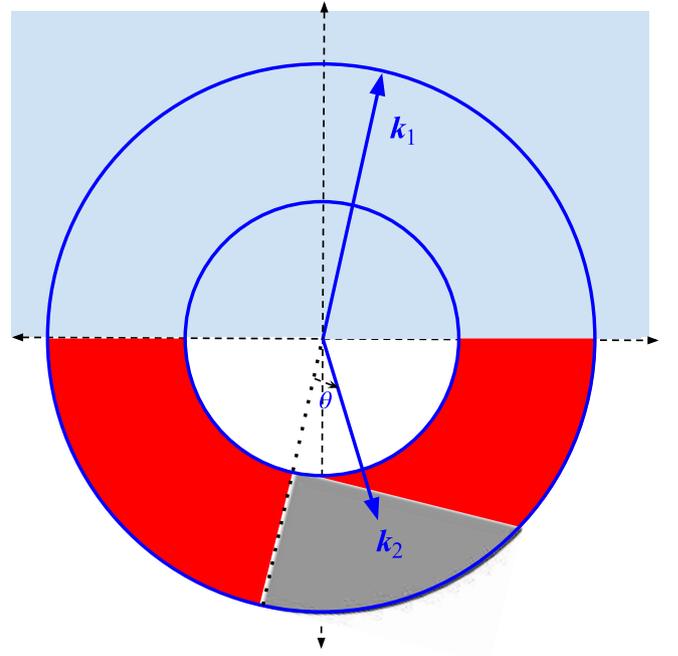}
\caption{An example of the method used to search $\kk_2$ for a $\kk_1$ on a 2D plane. The $\kk_2$ search is confined to the red shaded region, out of which the grey shaded region corresponds to all unique triangle configurations which satisfy Eq.~(\ref{eq:triangle4}).}
\label{fig:estimator}
\end{figure}

To avoid the aforementioned problems and to optimise the sampling of all possible unique triangle, we developed a new direct SABS estimation code \textsc{DviSukta}\footnote{Available at: \url{https://github.com/rajeshmondal18/DviSukta}}. The code is parallelized and uses the following approach:
\begin{itemize}
\item It starts by reading the brightness temperature data, which could be in real space or redshift space but has to be gridded in physical coordinates (Mpc), and performing a 3D Fourier transform of it. Alternatively, it starts by reading the data already in Fourier space.

\item For parallelization over multiple threads, the data is divided into $N_{\rm threads}$ equal parts. Each part is sent to a separate compute thread.

\item It searches for all possible $\kk_1$ and bins them. It uses equally spaced spherical logarithmic binning for $k_1$. However, the binning scheme can be easily changed.

\item Under the $\kk_1$ loop, it searches for all possible $\kk_2$. For this it
makes use of Eq.~(\ref{eq:triangle3}) and partial use of Eq.~(\ref{eq:triangle4}). Figure~\ref{fig:estimator} illustrates an example of this method for a $\kk_1$-vector in a 2D plane. In this particular example, the $\kk_2$ search is confined to the red shaded region, out of which the grey shaded region corresponds to all unique triangle configurations which satisfy Eq.~(\ref{eq:triangle4}). This trick drastically reduces the search from $N_{\rm G}^3$ to roughly $k_1^3V/(4\pi)^2$. Note that the algorithm also takes care of all possible orientations (i.e. all rigid body rotations) of a triangle in this search.
\item It maps the SABS values from $(k_2, k_3)$ space to $(n, \cos{\theta})$ space, and bins them. It uses equally spaced linear bins for $n$ and $\cos{\theta}$. However, also here the choice of binning is very flexible. 

\item It waits for all threads to complete and joins them.

\item It performs the bin averaging and produces  $\bar{B}(k_1,n,\cos{\theta})$.
\end{itemize}
The speed of this algorithm obviously depends on the number of threads used. To give an indication we measured the speed using an Intel core i7 dual-core laptop with 4 threads (roughly 80\% CPU utilisation). Using the four threads we found that the SABS estimation for a $128^3$ data set takes approximately 25 minutes. We have used 10 equally spaced logarithmic bins for the $k_1$ range and 10 linear bins for both $n$ and $\cos{\theta}$ range. However, as most of the computing time is used for finding the unique triangles, these numbers hardly influence the execution time. For different mesh sizes the computing time was found to scale approximately as $N_\mathrm{G}^{4.8}$. The optimisations outlined above clearly lead to a considerable improvement over the standard scaling for direct methods,  $\propto N_\mathrm{G}^6/2$.


\section{Simulating the EoR 21-cm Signal}
\label{sec:sim}
This work uses the same LC EoR 21-cm signal as was used in \citet{mondal18, mondal19}, to which we refer for a detailed description of the simulation methodology. For the benefit of the reader, we provide a brief description of the semi-numerical technique used for simulating the coeval signal and summarise how we have generated the redshifted 21-cm LC signal. Note that we work with an inside-out reionization model in the sense that we assume the collapsed dark matter halos host the ionizing sources, and the distribution of the hydrogen gas follows the underlying dark matter field. Our semi-numerical technique employs the excursion-set formalism of \citealt{furlanetto04b}) and the homogeneous recombination scheme of \citet{choudhury09b} to produce ionization maps at a given redshift.

Our procedure of generating the coeval ionization maps consists of three major steps. First, a particle-mesh (PM) based $N$-body code\footnote{Available at: \url{https://github.com/rajeshmondal18/N-body}} is used to generate dark matter distributions at the desired redshifts. Here the matter distributions were simulated within a comoving volume of $[300.16\,{\rm Mpc}]^3$ using $4288^3$ grids (which corresponds to $70\,{\rm kpc}$ grid spacing). We have used $2144^3$ dark matter particles that corresponds to a mass resolution $1.09 \times 10^8\,M_{\sun}$. In the next step, a Friends-of-Friends (FoF) halo finder\footnote{Available at: \url{https://github.com/rajeshmondal18/}FoF-Halo-finder} algorithm is used to find the collapsed objects within the dark matter distributions. We set the linking-length parameter at $0.2$ times the mean inter-particle separation. We consider a group of particles to be a halo if it consists of at least $10$ dark matter particles\footnote{Resolving halos with 10 particles is not very realistic, particularly for a PM $N$-body code, where a group of minimum $\sim 50$ particles are generally used to reliably identify a cluster or halo. However, in our simulations, we find that the halo mass function obtained using a minimum group size of 10 particles for halo identification is in good agreement \citep{majumdar12, Das2018} with the theoretical mass function of \citet{sheth02} with the fitting function adopted from \citet{jenkins01}.}, leading to a minimal halo mass of $M_{\rm min} = 1.09 \times 10^9\,M_{\sun}$.

Lastly, we generate coeval ionization cubes using our semi-numerical code {\sc ReionYuga}\footnote{Available at: \url{https://github.com/rajeshmondal18/ReionYuga}}. It follows a prescription where the number of ionizing photons $N_\gamma$ produced by a source is directly proportional to its host halo mass $M_{\rm h}$ (where $M_{\rm h} \ge M_{\rm min}$). We can write this prescription in the form (see eq. 3 of \citealt{majumdar14})
\begin{equation}
N_\gamma(M_{\rm h}) = N_{\rm ion} \frac{\Omega_{\rm b}}{\Omega_{\rm m}} \frac{M_{\rm h}}{m_{\rm H}}~,
\label{eq:nion}
\end{equation}
where $N_{\rm ion}$ is the dimensionless proportionality constant, ($\Omega_{\rm b}$, $\Omega_{\rm m}$) are the cosmological density parameters respective to the baryons and matter, and $m_{\rm H}$ is the mass of a hydrogen atom. $N_{\rm ion}$ therefore characterises the efficiency of sources and is one of the parameters in our simulation. $M_{\rm min}$ is another parameter in our simulations which is kept at the fiducial value of $1.09\times 10^9\,M_{\sun}$. The mean free path $R_{\rm mfp}$ of the ionizing photons in the ionized IGM is the third parameter. One can achieve different reionization histories by varying these three parameters. We, however, set $N_{\rm ion}$ and $R_{\rm mfp}$ to $23.21$ and $20\,{\rm Mpc}$, respectively. This results in $50\%$ ionization by $z = 8$ and complete ionization by $z\approx 6$. We obtain a reionization history which is consistent with \citet{Davies2018}, and the corresponding Thomson scattering optical depth $\tau=0.056$ agrees with measured CMB optical depth (see e.g. \citealt{planck_tau16}).

Finally, the \HI density field in the coeval cubes is represented by \HI particles. The total number of \HI particles, as well as the positions and peculiar velocities of each particle are the same as in the $N$-body simulation. The mass of each \HI particle is calculated by interpolating the neutral hydrogen fraction $\xHI$ from its eight adjacent grid points.

\begin{figure}
\centering
\includegraphics[width=0.48\textwidth, angle=0]{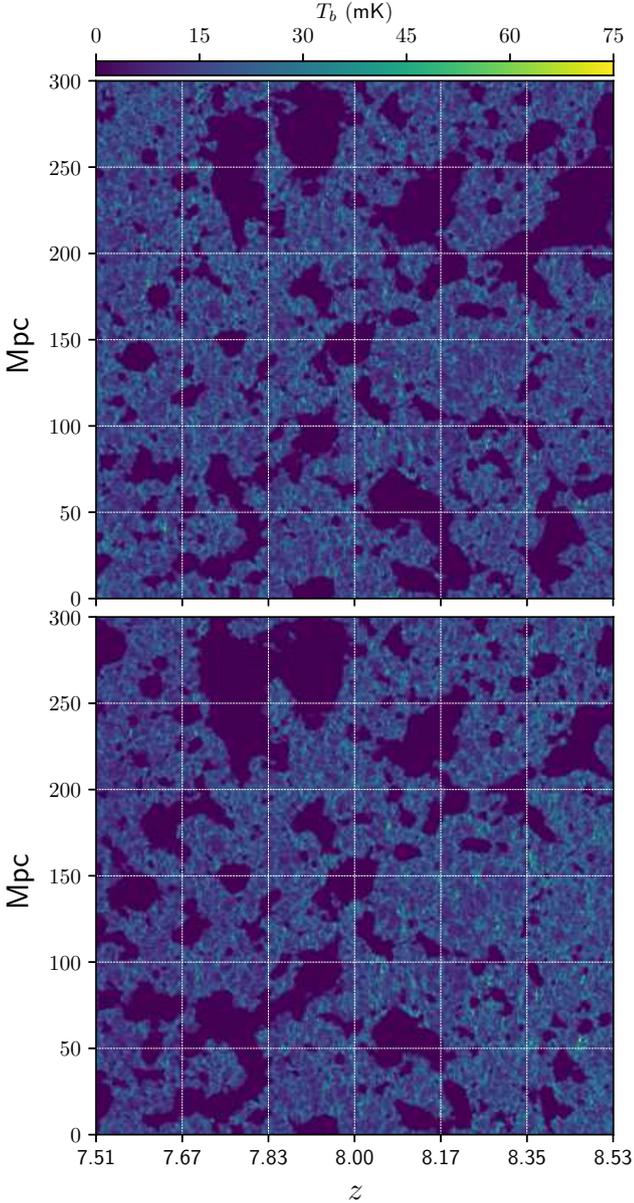}
\caption{The EoR 21-cm brightness temperature maps for the coeval (top) and LC (bottom) simulations. For further details, see fig. 4 of \citet{mondal18}.}
\label{fig:HI_map}
\end{figure}

We have generated $25$ such coeval \HI particle cubes centred at different $z_i$ that span the redshift range $z= 7.5$ to $z = 8.53$. Our choice of $z$ range is such that the comoving depth of the corresponding LC volume corresponds to the size of our $N$-body simulation, $300.16\,{\rm Mpc}$. The cubes are at non-uniform $\Delta z$ intervals such that the difference in mean neutral hydrogen fraction $\Delta \xb$ is approximately constant between consecutive coeval ionisation cubes. Each redshift $z_i$ corresponds to a different comoving radial distance $r_i$ and vice-versa. Therefore, to construct the LC box, we take out the region between $r_i$ and $r_{i+1}$ from the corresponding coeval snapshot at the $z_i$ and stitch them sequentially. Note that the stitching is performed in real space onto which the redshift-space distortions are applied to generate the final EoR 21-cm LC signal. We follow the prescription presented in \citet{majumdar13} to map the \HI particle distribution to the redshift space. Figure~\ref{fig:HI_map} shows the redshifted EoR 21-cm brightness temperature maps for the same section through the coeval and LC simulations centred at $z=8$.

In addition to the LC volume, we have simulated $50$ statistically independent realizations of the coeval volumes at $z=8$. This ensemble of redshifted 21-cm signals is used to estimate the CV errors in the SABS.

It is important to note that the results presented throughout this work are based on a single reionization history. Different reionization scenarios (i.e. slower or faster) can yield different amplitudes for the LC effect at different length scales. Therefore, our results cannot be considered to be a general feature but rather to provide a model specific demonstration of the LC effect on the SABS. 


\section{Results}
\label{sec:result}

\begin{figure}
\psfrag{pk}[c][c][1.1][0]{${\Delta^2(k)}~[{\rm mK}]^2$}
\psfrag{diff}[c][c][.9][0]{${(\Delta^2_{\rm LC} - \Delta^2_{\rm C})}/\Delta^2_{\rm C}$}
\psfrag{k}[c][c][1.1][0]{$k~({\rm Mpc}^{-1})$}
\psfrag{150Mpc}[c][c][1.1]{150 Mpc}
\psfrag{200Mpc}[c][c][1.1]{200 Mpc}
\psfrag{250Mpc}[c][c][1.1]{250 Mpc}
\psfrag{300Mpc}[c][c][1.1]{300 Mpc}
\psfrag{Coeval}[c][c][1.1]{~~~~Coeval}
\psfrag{LC}[c][c][1.1]{~~LC}
\centering
\includegraphics[width=.48\textwidth, angle=0]{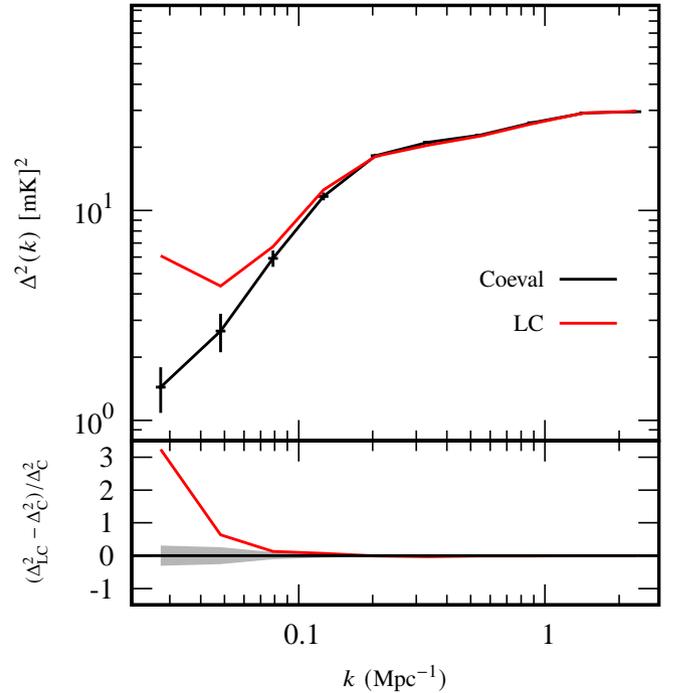}
\caption{The scale independent 21-cm SAPS $\Delta^2 (k)$ and the $1\sigma$ CV errors for the coeval signal (calculated using 50 statistically independent realizations). We also show the relative difference between $\Delta^2_{\rm LC}$ and $\Delta^2_{\rm C}$ where the grey shaded regions represent the $1\sigma$ CV errors.}  
\label{fig:pk}
\end{figure}

We expect the LC effect to be more pronounced and important when the mean hydrogen neutral fraction changes substantially over the observed bandwidth. The effect can thus be important over a relatively small bandwidth if this change is rapid, whereas it requires observing over a large bandwidth if reionization proceeds slower. If one accepts anisotropic sampling in $k$-space, the LC effect can be minimised by appropriately choosing the bandwidth \citep{datta14}. However, there is additional information in large bandwidth data which would then be discarded. Rather than considering it to be a problem and avoiding the LC effect, it can actually be included in the analysis. This is for example the approach taken by {\sc 21CMMC} (see e.g. \citealt{Greig2018}). Therefore, we consider the entire bandwidth of our simulated volumes to maximise the effect. We test this by first calculating the scale independent spherically averaged power spectrum (SAPS) $\Delta^2(k) = k^3 P(k)/2\pi^2$. In Fig.~\ref{fig:pk}, we show the SAPS for coeval ($\Delta^2_{\rm C}$) and LC ($\Delta^2_{\rm LC}$) simulations where the central redshift corresponds to $\xb = 0.5$. At this stage of reionization the large scale fluctuations typically are at a maximum. We also show the 1$\sigma$ CV errors calculated using the 50 statistically independent realizations of the coeval simulation. The CV errors scale as $1/\sqrt{V}$ \citep{Peacock1992} if the survey volume is increased while keeping the resolution and binning scheme the same. Therefore, one can predict CV errors (within the $k$-range shown in Fig.~\ref{fig:pk}) for any volume by scaling our predictions to account for the $1/\sqrt{V}$ dependence (see e.g. equation~31 of \citealt{mondal16}). We find that the LC effect is significant on length-scales $k \la 0.08\,{\rm Mpc}^{-1}$ and it introduces more than $50\%$ enhancement at scales $k \la 0.05\,{\rm Mpc}^{-1}$ reaching $\sim 200$ percent enhancement at our smallest $k$ values. Therefore, one should take the LC effects into account while making predictions using the SAPS. 


\begin{figure}
\psfrag{k}[c][c][1.1]{$k_1~({\rm Mpc}^{-1})$}
\psfrag{diff}[c][c][.9]{$(\Delta^3_{\rm LC} - \Delta^3_{\rm C})/\Delta^3_{\rm C}$}
\psfrag{bispec}[c][c][1.2]{$\Delta^3(k_1,n,\cos{\theta})~[{\rm mK}]^3$}
\psfrag{150Mpc}[c][c][1.1]{150 Mpc}
\psfrag{200Mpc}[c][c][1.1]{200 Mpc}
\psfrag{250Mpc}[c][c][1.1]{250 Mpc}
\psfrag{300Mpc}[c][c][1.1]{300 Mpc}
\psfrag{Coeval}[c][c][1.1]{~~~~Coeval}
\psfrag{LC}[c][c][1.1]{~~LC}
\centering
\includegraphics[width=.48\textwidth, angle=0]{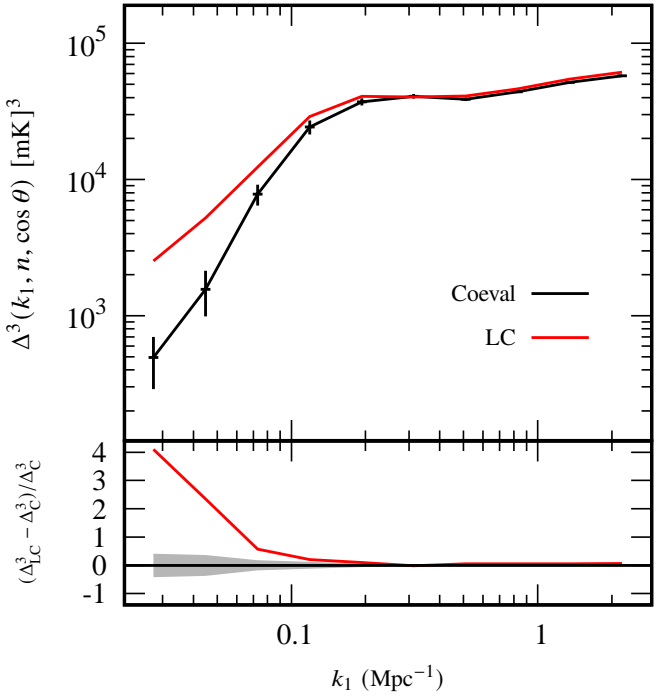}
\caption{The SABS for the limiting squeezed triangles ($n \xrightarrow{} 1$  i.e. $k_2 \xrightarrow{} k_1$ and $\cos{\theta} \xrightarrow{} 1$ i.e. $k_3 \xrightarrow{} 0$) and the $1\sigma$ CV errors for the coeval signal (calculated using 50 statistically independent realizations). We also show the relative difference between $\Delta^3_{\rm LC}$ and $\Delta^3_{\rm C}$ where the grey shaded regions represent the $1\sigma$ CV errors.}  
\label{fig:SABS_1-1}
\end{figure}

Now, we focus on the theme of this work namely the impact of the LC effect on the SABS. We do not discuss the entire theoretical background for the interpretation of 21-cm SABS from the EoR for which we refer to the discussion of figures~3 and~4 in \citet{majumdar20}. We present our results in terms of the scale independent SABS defined as
\begin{equation}
\Delta^3(k_1,n,\cos{\theta}) \equiv \frac{k_1^6\,n^3\, \bar{B}(k_1,n,\cos{\theta})}{(2\pi^2)^2} \, .
\end{equation}
Figure~\ref{fig:SABS_1-1} shows this $\Delta^3(k_1,n,\cos{\theta})$ for the squeezed limit triangles ($n \xrightarrow{} 1$  i.e. $k_2 \xrightarrow{} k_1$ and $\cos{\theta} \xrightarrow{} 1$ i.e. $k_3 \xrightarrow{} 0$) for the LC and coeval simulations. The $1\sigma$ CV errors are shown for the coeval simulations. There exists a correspondence between the SAPS and the squeezed limit SABS \citep[see][and references therein]{2019JCAP...02..058G} and it is therefore not so surprising that we find somewhat similar results when comparing Figs.~\ref{fig:pk} and~\ref{fig:SABS_1-1}. However, a close inspection reveals two important differences. First of all the CV error is larger for the SABS as compared to SAPS, even though  the number of measurements in a $k$-bin for SABS is larger than for SAPS. This is due to the fact that the SABS is a higher order statistics. Second, the LC effect is more significant for SABS. The LC effect is important on scales $k \la 0.1\,{\rm Mpc}^{-1}$. The enhancement due to LC effect is $25$~percent at $k_1 \approx 0.1\,{\rm Mpc}^{-1}$ and reaches more than $200$~percent at $k_1 \approx 0.05\,{\rm Mpc}^{-1}$ and even higher for our smallest $k_1$ values. Therefore, the LC effect has a larger impact on a measurement of the SABS, at least for the squeezed limit triangles.

\begin{figure}
\psfrag{k}[c][c][1.2]{$k_1~({\rm Mpc}^{-1})$}
\psfrag{diff}[c][c][1.2]{$(\Delta^3_{\rm LC} - \Delta^3_{\rm C})/\Delta^3_{\rm C}$}
\psfrag{B}[c][c][1.2]{$\Delta^3(k_1,n,\cos{\theta})~[{\rm mK}]^3$}
\psfrag{Coeval}[c][c][1.2]{~~~~Coeval}
\psfrag{LC}[c][c][1.2]{~~~~LC}
\centering
\includegraphics[width=0.48\textwidth, angle=0]{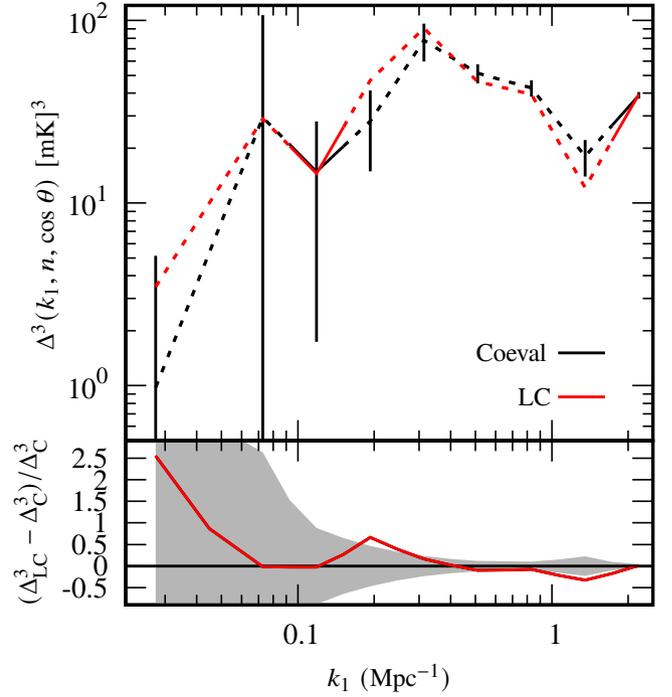}
\caption{As Fig.~\ref {fig:SABS_1-1} but for the limiting equilateral triangles ($n \xrightarrow{} 1$  i.e. $k_2 \xrightarrow{} k_1$ and $\cos{\theta} \xrightarrow{} 0.5$ i.e. $k_3 \xrightarrow{} k_1$). The solid lines and dashed lines represent positive and negative values of the SABS, respectively.}  
\label{fig:SABS_0.5-1}
\end{figure}

Figure~\ref{fig:SABS_0.5-1} shows the SABS for equilateral triangles ($n \xrightarrow{} 1$  i.e. $k_2 \xrightarrow{} k_1$ and $\cos{\theta} \xrightarrow{} 0.5$ i.e. $k_3 \xrightarrow{} k_1$). As shown by \citet{majumdar18}, the bispectrum for equilateral triangles oscillates between negative to positive as function of $k_1$ for a toy model with a fixed bubble radius, and peaks around the scale corresponding to that characteristic bubble radius. However, for a model with a log-normal bubble size distribution the SABS resembles more a power law, and has a transition from negative to positive at a very small scale. We expect the EoR 21-cm signal to be somewhere in between these two cases. For both LC and coeval simulations, the SABS peaks around $k = 0.31\,{\rm Mpc}^{-1}$, which roughly corresponds to a characteristic bubble size of $\sim 20\,{\rm Mpc}$. However, in terms of the LC effect we find that the changes it introduces are smaller than the CV errors and therefore not very significant.

\begin{figure*}
\psfrag{cost}[c][c][1.2]{$\cos{\theta}$}
\psfrag{k=0.045}[c][c][1.2]{~~~~~~~~$k_1 = 0.045\,{\rm Mpc}^{-1}$}
\psfrag{k=0.19}[c][c][1.2]{~~~~~~~~$k_1 = 0.19\,{\rm Mpc}^{-1}$}
\psfrag{k=1.35}[c][c][1.2]{~~~~~~~~$k_1 = 1.35\,{\rm Mpc}^{-1}$}
\psfrag{B}[c][c][1.2]{$\Delta^3(k_1,n,\cos{\theta})~[{\rm mK}]^3$}
\psfrag{Coeval}[c][c][1.2]{Coeval}
\psfrag{LC}[c][c][1.2]{LC}
\psfrag{L-isosceles}[c][c][1.2]{L-isosceles~~~~~~~}
\centering
\includegraphics[width=1.\textwidth, angle=0]{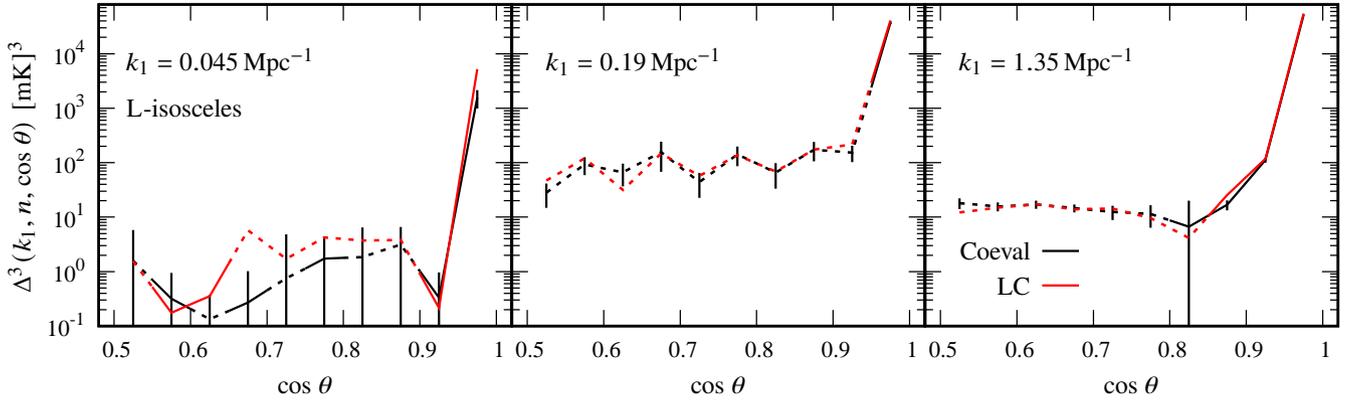}
\caption{The SABS for the limiting L-isosceles ($n \xrightarrow{} 1$ i.e. $k_2 \xrightarrow{} k_1$) and the $1\sigma$ CV errors for the coeval signal at $k_1 = 0.045,\, 0.19,\, 1.35\,{\rm Mpc}^{-1}$. The solid lines and dashed lines represent positive and negative values of the SABS, respectively.}  
\label{fig:SABS_n1}
\end{figure*}

Figure~\ref{fig:SABS_n1} shows the SABS for the L-isosceles triangles ($n \xrightarrow{} 1$ i.e. $k_2 \xrightarrow{} k_1$) as a function of $\cos{\theta}$ at three different scales $k_1 = 0.045,\, 0.19,\, 1.35\,{\rm Mpc}^{-1}$ for the coeval and LC results. It also includes the $1\sigma$ CV errors, which are calculated from the 50 statistically independent realisations of the coeval simulation. In this case, the length of the smallest arm (and the area) of the triangles decreases with the value of $\cos{\theta}$. We expect the SABS to peak for the squeezed limit ($\cos{\theta} \xrightarrow{} 1$) triangle configuration. Indeed, we see that the magnitude of the SABS is highest around the $\cos{\theta} = 0.975$ bin and for all values of $k_1$ falls very sharply (two orders of magnitude) for smaller values of $\cos{\theta}$. We also notice that for these lower values of  $\cos{\theta}$, the values are nearly independent of $\cos{\theta}$. We further see that the magnitude of the SABS overall increases when we move from large scales to small scales until $k \sim 0.2\,{\rm Mpc}^{-1}$, which is roughly the characteristics bubble size. For larger $k$ the magnitude of the SABS remains more or less the same which is qualitatively similar to what was seen for the squeezed triangles (Fig.~\ref{fig:SABS_1-1}). At large scales ($k_1 = 0.045\,{\rm Mpc}^{-1}$, left panel of Fig.~\ref{fig:SABS_n1}) we find that the amplitude of the SABS is small ($\sim 1$) and the values oscillate between positive and negative. However, the latter behaviour can be attributed to the large CV error associated the SABS at these scales. The LC effect falls mostly within the CV errors except for $\cos{\theta} \xrightarrow{} 0.675$ (i.e. $k_3 \xrightarrow{} 4k_1/5$) and for the squeezed limit. It only has a negligible impact on scales $k \ge 0.19\,{\rm Mpc}^{-1}$.

\begin{figure*}
\psfrag{n}[c][c][1.2]{$n$}
\psfrag{k=0.045}[c][c][1.2]{~~~~~~~~$k_1 = 0.045\,{\rm Mpc}^{-1}$}
\psfrag{k=0.19}[c][c][1.2]{~~~~~~~~$k_1 = 0.19\,{\rm Mpc}^{-1}$}
\psfrag{k=1.35}[c][c][1.2]{~~~~~~~~$k_1 = 1.35\,{\rm Mpc}^{-1}$}
\psfrag{B}[c][c][1.2]{$\Delta^3(k_1,n,\cos{\theta})~[{\rm mK}]^3$}
\psfrag{Coeval}[c][c][1.2]{Coeval}
\psfrag{LC}[c][c][1.2]{LC}
\psfrag{Linear triangles}[c][c][1.2]{Linear triangles~~~~~~}
\centering
\includegraphics[width=1.\textwidth, angle=0]{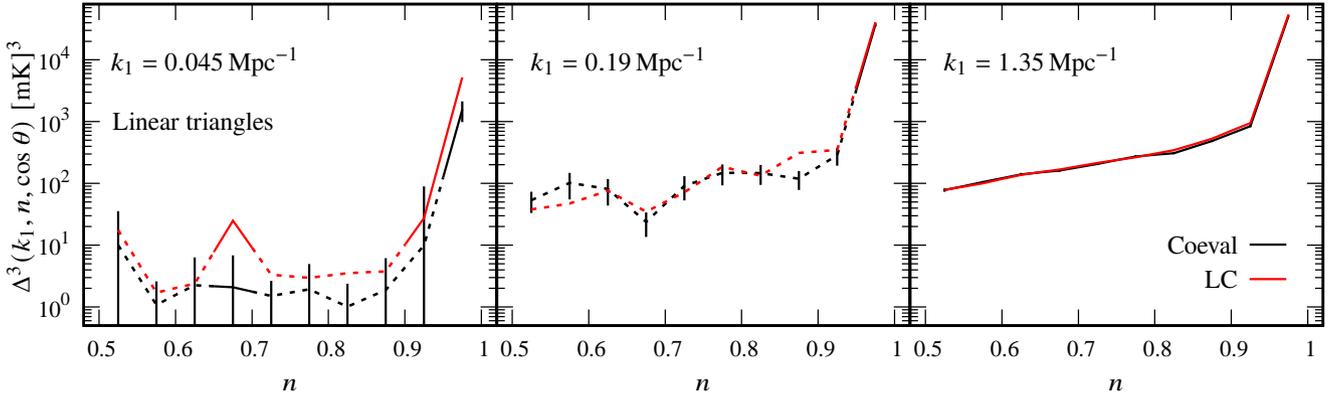}
\caption{As Fig.~\ref{fig:SABS_n1} but for the limiting linear triangles ($\cos{\theta} \xrightarrow{} 1$).}  
\label{fig:SABS_cost1}
\end{figure*}

Lastly we consider the case of the linear triangles ($\cos{\theta} \xrightarrow{} 1$). In this case, the length of the second largest arm $k_2$ of the triangles increases with the value of $n$ and so we show the results for our three different $k$ values as a function of $n$. Figure~\ref{fig:SABS_cost1} shows that the magnitudes found are similar as for the L-isosceles (Fig.~\ref{fig:SABS_n1}). However, there are a few differences. The magnitude of the SABS for the linear triangles slowly increases as $n$ increases. The results of the L-isosceles and linear triangles also show sign differences. These are due to the causes discussed in the next paragraph. Regarding the LC effect, we can draw a similar conclusion as above, namely that it really only exceeds the CV errors for $k_1 =0.045\,{\rm Mpc}^{-1}$ and even there only significantly for the squeezed limit triangles ($n\xrightarrow{}1$).

\begin{figure*}
\centering
\psfrag{nn}[c][c][1.2]{$n$}
\psfrag{cost}[c][c][1.2]{$\cos{\theta}$}
\psfrag{B}[c][c][1.2]{$\Delta^3(k_1,n,\cos{\theta})~[{\rm mK}]^3$}
\psfrag{k=0.045}[c][c][1.2]{$k_1 = 0.045~{\rm Mpc}^{-1}$}
\psfrag{k=0.19}[c][c][1.2]{$k_1 = 0.19~{\rm Mpc}^{-1}$}
\psfrag{k=1.35}[c][c][1.2]{$k_1 = 1.35~{\rm Mpc}^{-1}$}
\psfrag{R}[c][c][1.]{$\Delta^3_{\rm LC}/\Delta^3_{\rm C}$}
\psfrag{Coeval}[c][c][1.]{Coeval\hspace{3.2cm}}
\psfrag{LC}[c][c][1.]{LC\hspace{3.2cm}}
\includegraphics[width=1.\textwidth, angle=0,scale=1]{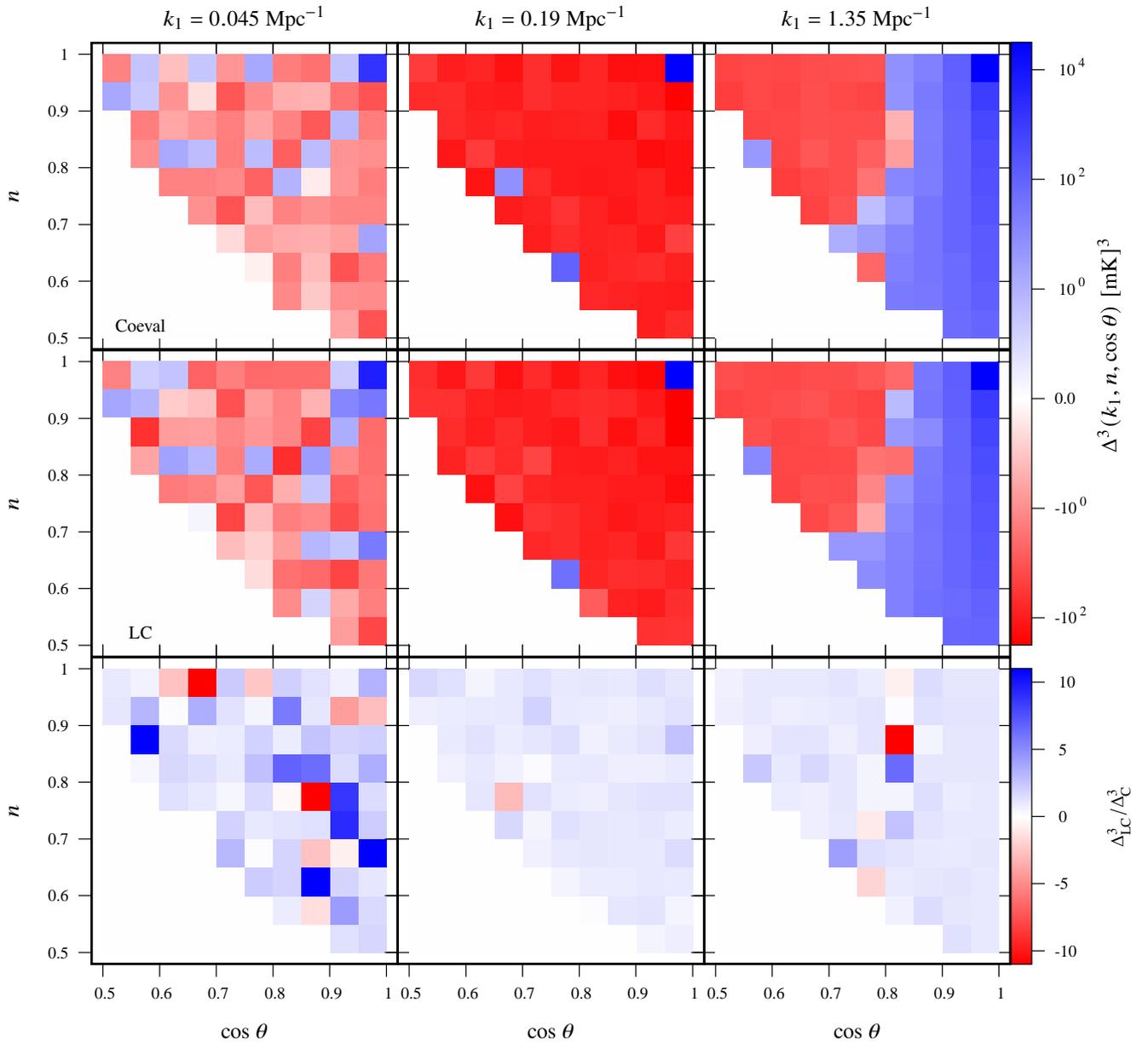}
\caption{The SABS for all unique triangle configurations for coeval (top row) and LC (middle row) simulations at $k_1 = 0.045,\, 0.19,\, 1.35\,{\rm Mpc}^{-1}$. The bottom row shows the ratio between $\Delta^3_{\rm LC}$ and $\Delta^3_{\rm C}$.}
\label{fig:SABS_matrix}
\end{figure*}

Figure~\ref{fig:SABS_matrix} shows the overview of the SABS for all unique triangle configurations for coeval and LC simulations at our three standard $k_1$ values of $0.045,\, 0.19$, and $1.35\,{\rm Mpc}^{-1}$. In the bottom panels, we show the ratio between the LC SABS ($\Delta^3_{\rm LC}$) and the coeval SABS ($\Delta^3_{\rm C}$). This representation of all bispectra for all unique triangles was introduced in \citet{bharadwaj20}. For a fixed $\cos{\theta}$ value, the length of the second largest side $k_2$ increases with $n$. While for a fixed $n$ value, the length of the shortest side $k_3$ decreases with $\cos{\theta}$. The different aspects of this figure can be understood in the following way: the EoR 21-cm signal $\delta_{\rm 21\,cm}$ is a multiplication of hydrogen density field $\delta_{\rm H}$ and the neutral fraction field $\delta_{\xHI}$. The bispectrum for the $\delta_{\rm H}$ field is always positive. Therefore, the EoR 21-cm SABS becomes negative due to the presence of $\delta_{\xHI}$, as the inside-out reionization implemented in our simulations implies $\delta_{\xHI}$ and $\delta_{\rm H}$ are anti-correlated. However, this anti-correlation is scale dependent (see e.g. figure~2 bottom panel of \citealt{lidz07}), they are perfectly anti-correlated on large scales, while the anti-correlation becomes weak at smaller scales. Therefore, the $k$ range can roughly be divided into two regions, one which is substantially larger than the typical size of ionized regions ($k \la 0.2\,{\rm Mpc}^{-1}$), and another which is substantially smaller than this ($k \ga 0.4\,{\rm Mpc}^{-1}$). Simply put, depending on the different combinations of the three $k$ modes, we can have negative SABS (e.g. all three $k$'s are small, or $k_1$ is large and $k_2$, $k_3$ are small, etc.) and positive SABS (e.g. all three $k$'s are large, or $k_1$, $k_2$ are large and $k_3$ is small, etc.). However, for a rigorous understanding of all the features, one would need to do a full decomposition analysis of the SABS, similar to what was done for the SAPS in \citet{lidz07}. Here, we do not discuss this further and focus on the LC effect. For a detailed discussion on this point, the reader is referred to section~5 of \citet{majumdar20}.

This is the first study on the redshifted 21-cm bispectrum which properly takes the CV errors into account. Above in Figs.~\ref{fig:SABS_1-1}, \ref{fig:SABS_0.5-1}, \ref{fig:SABS_n1} and \ref{fig:SABS_cost1} we included these errors and saw that they are non-negligible for many of the triangle configurations and $k_1$ values. Figure~\ref{fig:SABS_matrix} does however not show the CV errors and therefore does not allow us to properly assess the impact of the LC effect. Below in Section~\ref{sec:sensitivity} we will consider the combined effects of CV errors and instrumental noise to derive signal-to-noise ratio\,(SNR) for the SABS for all unique triangle configurations.

The SABS for all unique triangle configurations roughly follow a general trend. We see that the magnitude of the SABS increases with $k$ for $k \la 0.2\,{\rm Mpc}^{-1}$ as the non-Gaussianity increases with $k$ on these scale. At small length-scales $k \ga 0.4\,{\rm Mpc}^{-1}$, however, it saturates. Those length-scales are roughly below the characteristics bubble size. However, this does not necessarily mean that the non-Gaussianity on these length-scales is more or less constant. It is perfectly possible to have structure in the higher order polyspectra (e.g. trispectrum; \citealt{mondal16}) on these length-scales. As expected, we see the LC effect is important at large scales, exemplified by the case of $k_1 = 0.045\,{\rm Mpc}^{-1}$ (Fig.~\ref{fig:SABS_matrix}, bottom left panel), although no clear pattern can be discerned. Interestingly at $k_1 = 1.35\,{\rm Mpc}^{-1}$ (bottom right panel) around $\cos{\theta}\approx 0.8$ we see two cases which appear to show a strong impact of the LC effect. However, inspection of the CV errors reveal these to not statistically significant (see e.g. right hand panel of Fig.~\ref{fig:SABS_n1}).


\section{The sensitivity to the SABS for SKA-Low}
\label{sec:sensitivity}
Here we consider the detectability of the EoR 21-cm SABS in future observations with SKA-Low. However, the methodology presented could of course be applied to any other radio-interferometer. We would like to start by pointing out that all previous studies presenting error predictions for the SABS \citep[see e.g.][]{yoshiura2015, watkinson19, ma21, watkinson21} have assumed the errors to behave as if the observed signal was a Gaussian random field. This assumption, however, is counter-intuitive and also under-predicts the CV errors (e.g. \citealt{mondal15, mondal16, mondal17, shaw19, shaw20}). As a result, previous SABS sensitivity estimates predict an unrealistically large SNR on large scales (small $k$ bins) where the cosmic variance dominates. We avoid making this assumption and compute the exact variance numerically using a signal ensemble. It consists of $50$ statistically independent realizations of coeval signal that is a sum of the cosmological 21-cm signal and the Gaussian system noise. We consider an optimistic scenario where only Gaussian system noise contaminates the signal, and assume the signal is free from foregrounds and other systematic errors. Note that the sensitivity predictions presented here are based on a single reionization history. A different reionization scenario or a different amplitude of the signal may yield different sensitivity predictions. Therefore, to understand the impact of the light cone effect in a robust manner one would need to consider various possible reionization models and histories. This work is a preliminary step in that direction. Therefore the results presented here are indicative of how system noise and CV errors would affect the measurement of the SABS. We next discuss how we generate the observed signal ensemble.

For a radio interferometric array, the primary observables are the visibilities. These are recorded at baselines $\mathbfit{U}=\mathbfit{d}/\lambda_i$ and the corresponding frequency $\nu_i$ with $\mathbfit{d}$ being the antenna separation projected onto the sky plane. For our interferometer we use the current proposed SKA-Low configuration \citep{SKA_Low_v2} with $512$ stations, each having a diameter $D=35~{\rm m}$. We consider a mock observation where the instruments tracks a patch in the sky at ${\rm DEC}=-30^\circ$ for $8~{\rm hrs}$ per night with $60~{\rm secs}$ integration time\footnote{We do not take into account that the sensitivity of the proposed SKA-Low antennas depends on zenith distance. This will increase the noise levels and may also make it inefficient to track the same patch over such a long time.}. Following the steps in \citet{shaw19}, the baseline tracks are generated at the frequency corresponding to $z=8$ (see e.g. figure~8 of \citealt{mondal20}). These baselines are linearly related with the perpendicular component of the $\kk$ mode, \textit{i.e.} $\kk_\perp= (2 \pi \mathbfit{U})/r_{\rm c}$ where $r_{\rm c}$ is the comoving distance corresponding to the redshift. We grid the $\kk_\perp$ plane with $\Delta \kk_\perp = 2\pi/L$ which is the same gridding as we use for the 21-cm signal simulations cube with size $L$. The baselines are then associated to the nearest grid points $\kk_{\rm G}$ to obtain the baseline sampling function $\tau(\kk_{\rm G})$. Note that the results in this section do not include the $k_{\perp} = 0$ modes as these modes are not measurable by interferometric experiments. These modes carry the information of the variation of the mean (global) signal along the LoS in the LC volume. As a result the LC effect on the SAPS for our $18\,{\rm MHz}$ simulation becomes statistically insignificant.

The baseline distribution changes along the LoS direction as a function of the observing frequency. However the change in baselines will be a few percent for the frequency interval considered here, and we ignore this in our analysis. The 3D Fourier volume is then filled by using the same gridded baseline distribution along the entire $k_\parallel$ axis. 

With this gridded baseline distribution in place, we generate the system noise visibilities at every grid point $\kk_{\rm G}$ using
\begin{equation}
\Delta_{\rm N}(\kk_{\rm G})=\sqrt{\frac{VP_{\rm N}(\kk_{\rm G})}{2}} ~[a(\kk_{\rm G}) + i b(\kk_{\rm G})]~,
\end{equation}
where $a(\kk_{\rm G})$ and $b(\kk_{\rm G})$ are two Gaussian random variables with zero mean and unit variance. $P_{\rm N}(\kk_{\rm G})$ is the system noise power spectrum which we compute at every grid point $\kk_{\rm G}$ following equation~1 of \citet{shaw19}. Details of the noise power spectrum computations can be found in section~3 of \citet{shaw19}. Note that this analysis avoids the Fourier cells which are not sampled by the baseline tracks ($\tau(\kk_{\rm G})=0$). In order to generate an ensemble for the observed signal, we simulate $50$ statistically independent realizations of the system noise map within the same coeval volume of the 21-cm signal, and add the system noise and 21-cm signal. We finally estimate the mean SABS and the errors directly from this ensemble without any approximation. Therefore, these error estimates have contributions from both the CV and the system noise. Note that, unlike for the SAPS, the SABS estimates (by default) are free from the noise bias as the system noise is Gaussian.

\begin{figure}
\psfrag{k}[c][c][1.2]{$k_1~({\rm Mpc}^{-1})$}
\psfrag{diff}[c][c][1.2]{$(\Delta^3_{\rm LC} - \Delta^3_{\rm C})/\Delta^3_{\rm C}$}
\psfrag{bispec}[c][c][1.2]{$\Delta^3(k_1,n,\cos{\theta})~[{\rm mK}]^3$}
\psfrag{300Mpc}[c][c][1.2]{$\sim 300$\,Mpc}
\psfrag{Coeval}[c][c][1.2]{Coeval}
\psfrag{LC}[c][c][1.2]{LC}
\psfrag{128hrs}[c][c][1.2]{128 hrs}
\psfrag{1024hrs}[c][c][1.2]{1024 hrs}
\centering
\includegraphics[width=.48\textwidth, angle=0]{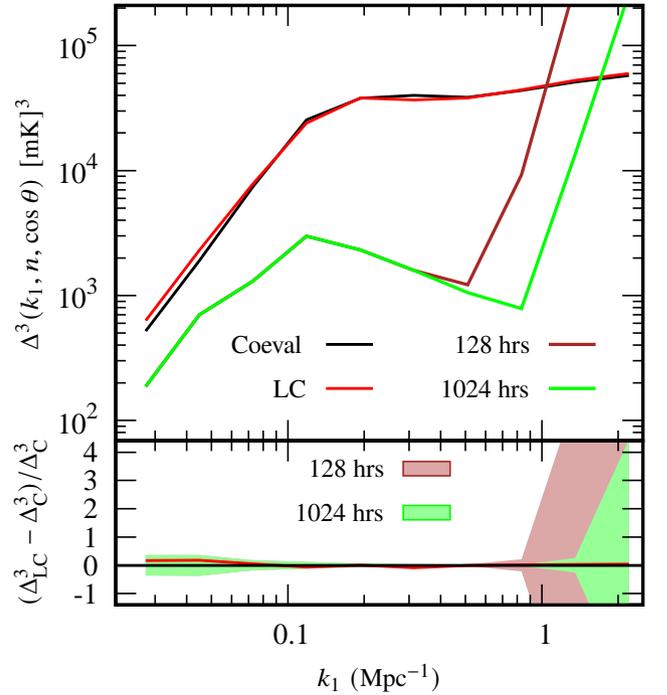}
\caption{The SABS for the limiting squeezed triangles ($n \xrightarrow{} 1$ i.e. $k_2 \xrightarrow{} k_1$ and $\cos{\theta} \xrightarrow{} 1$ i.e. $k_3 \xrightarrow{} 0$) and the corresponding $1\sigma$ error estimates for $t_{\rm obs} = 128$\,hrs and $1024$\,hrs. We also show the relative difference between $\Delta^3_{\rm LC}$ and $\Delta^3_{\rm C}$ where the brown and green shaded regions represent the $1\sigma$ errors for $t_{\rm obs} = 128$\,hrs and $1024$\,hrs, respectively.}  
\label{fig:SABS+noise_1-1}
\end{figure}

Figure~\ref{fig:SABS+noise_1-1} shows the SABS error estimates for the squeezed limit triangles. These errors are computed using the ensemble of  coeval signals for observation times $t_{\rm obs} = 128$\,hrs and $1024$\,hrs. In Section~\ref{sec:result}, we have discussed the CV only predictions which corresponds to a limiting case where system noise contributions $\xrightarrow{} 0$ as $t_{\rm obs} \xrightarrow{} \infty$. The first point to note here is that the system noise contribution to the observed SABS error is expected to scale as $t_{\rm obs}^{-3/2}$. One can verify this scaling by comparing the error estimates obtained for the two different observation times at the scales where the system noise dominates, i.e. $k_1 \ga 0.5\,{\rm Mpc}^{-1}$. In contrast to this and as expected, the SABS error estimate is dominated by the CV errors on large scales $k_1 \la 0.4\,{\rm Mpc}^{-1}$. We find the squeezed limit SABS is detectable on length-scales $k_1 \la 1\,{\rm Mpc}^{-1}$. Considering the bottom panel, we find that at large length-scales the system noise contribution does not make any considerable change in the total error budget as compared to the CV only case (Fig.~\ref{fig:SABS_1-1}). This is simply because the system noise contribution is negligible compared to the CV contribution on these scales. However, we also see that the removal of $k_{\perp} = 0$ modes severely diminishes the impact of the LC effect on the squeezed limit SABS.

The reason for choosing squeezed limit triangles for this plot is because of the high SNR achieved. This is due to the two factors -- (1) the SABS itself peaks near the squeezed limit and (2) the corresponding CV errors are also the smallest. The magnitude of the SABS falls sharply for the other triangle shapes. In addition the CV errors also increase for them, thus causing a drop in the SNR values. The prospects for detection become even worse after including the system noise contribution. The sensitivity predictions for triangles of all shapes are discussed below. In the subsequent results, we concentrate on triangles with sizes within the range $0.05\,{\rm Mpc}^{-1} \la k_1 \la 0.5\,{\rm Mpc}^{-1}$ which is the optimum range for measuring SABS using SKA-Low.

\begin{figure*}
\centering
\psfrag{nn}[c][c][1.2]{$n$}
\psfrag{cost}[c][c][1.2]{$\cos{\theta}$}
\psfrag{snr}[c][c][1.2]{${\rm SNR}$}
\psfrag{k=0.072}[c][c][1.2]{$k_1 = 0.072~{\rm Mpc}^{-1}$}
\psfrag{k=0.19}[c][c][1.2]{$k_1 = 0.19~{\rm Mpc}^{-1}$}
\psfrag{k=0.51}[c][c][1.2]{$k_1 = 0.51~{\rm Mpc}^{-1}$}
\includegraphics[width=1.0\textwidth, angle=0,scale=1]{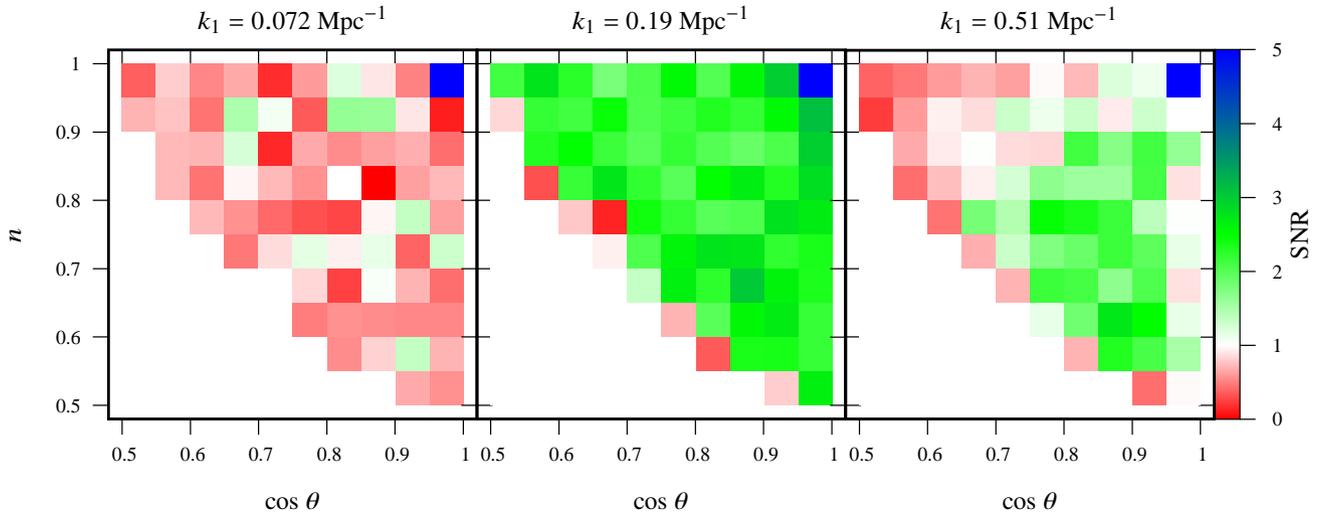}
\caption{The SNR for detecting the SABS for all unique triangle configurations for $t_{\rm obs}=1024$\,hrs for three different triangle sizes.}
\label{fig:SNR_matrix}
\end{figure*}

Figure~\ref{fig:SNR_matrix} shows the SNR predictions of SABS for all unique triangle configurations at $k_1 = 0.072,\, 0.19$, $0.51\,{\rm Mpc}^{-1}$ for $t_{\rm obs}=1024$\,hrs. The red, green and blue colours in the plot represent no detection, $\sim 3\sigma$ detection and $\ge 5\sigma$ detection respectively. We find that $\ge 5\sigma$ detection is only possible for the squeezed limit triangles. At $k_1 \la 0.072\,{\rm Mpc}^{-1}$, the SNR is largely governed by the CV errors and $\sim 2\sigma$ detection is possible for a few triangle configurations. A value of $k_1 \approx 0.19\,{\rm Mpc}^{-1}$ corresponds to the sweet spot between the CV errors and the system noise and we note that $\approx 3\sigma$ detection will be possible across almost the entire space of unique triangle configurations. Finally for $k_1 \ga 0.51\,{\rm Mpc}^{-1}$, the system noise contributions start dominating the total error estimate. At $k_1=0.51\,{\rm Mpc}^{-1}$, SKA-Low will be able to measure the SABS for obtuse triangles with $\approx 2\sigma$ confidence. For $k_1 > 0.51\,{\rm Mpc}^{-1}$ all triangles except the squeezed limit ones show ${\rm SNR} < 1$ and therefore we do not show the SNR plots for those scales.

\begin{figure*}
\psfrag{cost}[c][c][1.2]{$\cos{\theta}$}
\psfrag{k=0.072}[c][c][1.2]{\hspace{1.5cm}$k_1 = 0.072\,{\rm Mpc}^{-1}$}
\psfrag{k=0.19}[c][c][1.2]{\hspace{1.5cm}$k_1 = 0.19\,{\rm Mpc}^{-1}$}
\psfrag{k=0.51}[c][c][1.2]{\hspace{1.5cm}$k_1 = 0.51\,{\rm Mpc}^{-1}$}
\psfrag{B}[c][c][1.2]{$\Delta^3(k_1,n,\cos{\theta})~[{\rm mK}]^3$}
\psfrag{Coeval}[c][c][1.2]{Coeval}
\psfrag{LC}[c][c][1.2]{LC}
\psfrag{128hrs}[c][c][1.2]{128 hrs}
\psfrag{1024hrs}[c][c][1.2]{1024 hrs}
\centering
\includegraphics[width=1.\textwidth, angle=0]{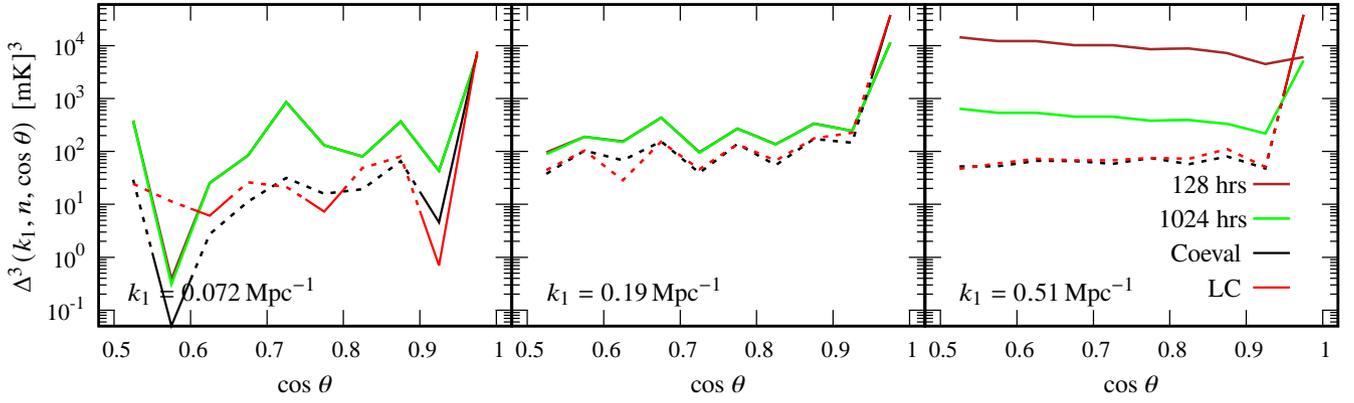}
\caption{The SABS for the limiting L-isosceles ($n \xrightarrow{} 1$ i.e. $k_2 \xrightarrow{} k_1$) at $k_1 = 0.072,\, 0.19$, $0.51\,{\rm Mpc}^{-1}$ and the corresponding $5\sigma$ rms error estimates for $t_{\rm obs}=128$\,hrs and $1024$\,hrs. The solid lines and dashed lines represent positive and negative values of the SABS, respectively. When only the green 1024~hrs line is shown, CV errors dominate over noise errors.}  
\label{fig:SABS+niose_n1}
\end{figure*}
\begin{figure*}
\psfrag{n}[c][c][1.2]{$n$}
\psfrag{k=0.072}[c][c][1.2]{\hspace{1.5cm}$k_1 = 0.072\,{\rm Mpc}^{-1}$}
\psfrag{k=0.19}[c][c][1.2]{\hspace{1.5cm}$k_1 = 0.19\,{\rm Mpc}^{-1}$}
\psfrag{k=0.51}[c][c][1.2]{\hspace{1.5cm}$k_1 = 0.51\,{\rm Mpc}^{-1}$}
\psfrag{B}[c][c][1.2]{$\Delta^3(k_1,n,\cos{\theta})~[{\rm mK}]^3$}
\psfrag{Coeval}[c][c][1.2]{Coeval}
\psfrag{LC}[c][c][1.2]{LC}
\psfrag{128hrs}[c][c][1.2]{128 hrs}
\psfrag{1024hrs}[c][c][1.2]{1024 hrs}
\centering
\includegraphics[width=1.\textwidth, angle=0]{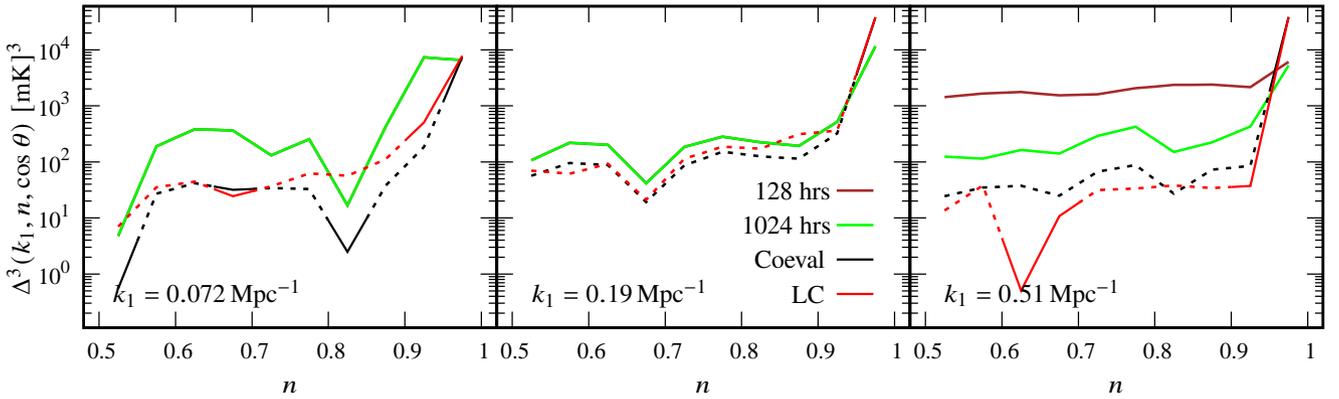}
\caption{Same as Fig.~\ref{fig:SABS+niose_n1} for the limiting linear triangles ($\cos{\theta} \xrightarrow{} 1$).}  
\label{fig:SABS+noise_cost1}
\end{figure*}

To better appreciate the impact of CV errors and noise on other bispectra than the squeezed limit ones, we show the SABS and the corresponding $5\sigma$ error estimates for the limiting L-isosceles and limiting linear triangles in Figures~\ref{fig:SABS+niose_n1} and~\ref{fig:SABS+noise_cost1}, respectively. For these two cases, the errors are very close to the magnitude of the SABS. Hence to clearly separate the curves, we set the confidence level to the high level of $5\sigma$. However, it is not uncommon to use this value as a benchmark to claim a detection in measurements. The results here are shown for the same $k_1$ values as in Figure~\ref{fig:SNR_matrix}, but for $t_{\rm obs}=128$\,hrs and $1024$\,hrs both. As evident from Fig.~\ref{fig:SABS_n1} and~\ref{fig:SABS_cost1}, the SABS for L-isosceles and linear triangles are dominated by CV errors on scales $k_1 \la 0.19\,{\rm Mpc}^{-1}$. As a consequence, we do not see any noticeable difference between $t_{\rm obs}=128$\,hrs (brown line) and $t_{\rm obs}=1024$\,hrs (green line) at $k_1 \le 0.19\,{\rm Mpc}^{-1}$, where the system noise contribution remains insignificant as compared to the CV errors. The LC effect is found to boost the SABS for a few shapes of the limiting L-isosceles and limiting linear triangles. Therefore, the SABS at $[n,\,\cos{\theta}] = [1,\,0.575]$ and $[0.825,\,1]$ for $k_1 = 0.072\,{\rm Mpc}^{-1}$ might be detectable. At $k_1=0.51\,{\rm Mpc}^{-1}$, however, the detectablility is restricted to the squeezed limit as the system noise prevails over the CV errors here. As a result, the difference between the error estimates for the 128~hrs and 1024~hrs observation times is visible over the entire $\cos\theta$ and $n$ ranges.


\section{Discussion and Conclusions}
\label{sec:conclusion}
In this paper we study the impact of the light-cone effect, cosmic variance and the expected noise level of SKA-Low on measurements of the 21-cm spherically averaged bispectrum. This study is based on a single reionization scenario for which we however generated 50 statistically independent realisations  of the coeval signal in order to estimate the CV errors. The scenario is simulated in a $[300\,{\rm Mpc}]^3$ comoving volume which roughly matches the size of the synthesised beam of SKA-Low for $z=8$.

For the calculation of the SABS we developed a new optimised direct estimation method, called \textsc{DviSukta}. It finds the values for all possible unique triangle configurations whilst avoids over-sampling/under-sampling at large/small scales by offering more flexible binning in the parameter space for all unique triangles ($k_1$, $n$, $\cos\theta$). Through optimised searching of parameter space this implementation of the direct method achieves an improved scaling with the number of grid points. Instead of the expected $N_\mathrm{\rm G}^{6}/2$ the computation time scales as $N_\mathrm{\rm G}^{4.8}$.

As previously found for the SAPS, we find that the LC effect affects the larger scales of the SABS. For most of the unique triangles the impact of the LC effect falls within the CV errors, with the notable exception of the squeezed limit triangles where LC effect is found to exceed the CV errors for scales $k \la 0.1\,{\rm Mpc}^{-1}$. Compared to the SAPS both the LC effect and the CV errors are found to be larger for these squeezed limit SABS.

We further calculate the detectability of the SABS for all unique triangles using up to 1024\,hrs of observing time with SKA-Low. For these predictions we do not include $k_{\perp} = 0$ modes as these modes are not measurable by interferometric experiments. When considering the impact of both CV errors and noise it is found that only the squeezed limit triangles can reach a SNR of more than 5 on length scales $k \la 1\,{\rm Mpc}^{-1}$. All other triangle configurations have lower SNR values. This is partly caused by the lower amplitude of these SABS and partly by their larger CV errors. In these SNR estimates we use the optimistic assumption that the observations can be calibrated to reach the theoretical noise level and that no systematic effects caused for example by residual foreground signals, remain.

Our results are based on a single scenario which reaches about 50 per cent reionization by redshift 8 and completes reionization around redshift 6. Such a scenario is consistent with existing observational constraints. However, these constraints still allow several other scenarios, including for example rather rapid scenarios \citep{Davies2018}, which would lead to a larger impact of the LC effect. In general scenarios with both larger and smaller LC effects can not yet be excluded.

We would like to point out that because of the non-Gaussian nature of the 21-cm signal, the CV errors cannot be reduced by combining different Fourier modes, i.e. by arbitrarily increasing your bin size \citep{mondal15}. However, one can still expect the CV errors in the estimated SABS to go down as $1/\sqrt{V}$ if the observational volume is increased while keeping the resolution and binning scheme the same \citep{mondal16}.

We have seen that the system noise errors mostly affect the larger $k_1$ values of the SABS and that the SNR for lower values is set by the CV errors. Unfortunately, as already shown by \citet{mondal15}, the CV errors cannot be easily estimated but have to be derived. As other scenarios can have different CV errors it is difficult to make definite statements about the impact of the CV errors. In other words, our results can only give an indication of how CV errors affect the measurement of the SABS. However, what is rather robust is that the squeezed limit SABS will always have the largest amplitude and therefore will be the version of the SABS which will have the largest SNR. Further studies of the bispectrum which only want to consider a limited number of triangle types should therefore at least consider the squeezed limit triangles as these will be easiest to measure.

One of the properties of the bispectrum for which it sometimes is criticised is the large number of different triangle configurations that can be selected which makes it a rather unwieldy statistical quantity which also is not easy to interpret. Our results indicate that even 1000\,hrs with SKA-Low will not yield a useful measurement of many of the possible triangle configurations and that it therefore may be best to focus on the squeezed limit triangles, thus simplifying the inherent complexity of the bispectrum. Furthermore, the squeezed limit bispectrum has a clear interpretation in terms of the position-dependent power spectra, as first pointed out in \citet{Chiang2014} and studied in the context of reionization by \citet{2019JCAP...02..058G}. One of the main reasons for measuring the bispectrum is to use it to break any degeneracy present in the SAPS \citep{mondal20b}. In view of our results it would be very useful to repeat the study in \citet{watkinson21} for squeezed limit triangles and test how well it performs in distinguishing different scenarios and setting constraints on model parameters.


\section*{Acknowledgements}
RM is supported by the Wenner-Gren Postdoctoral Fellowship. The authors would like to thank Prof. Somnath Bharadwaj for his useful discussion and insightful comments. A significant part of the bispectrum estimation were done using the computing resources available to the Cosmology with Statistical Inference (CSI) research group at IIT Indore. The authors would like to acknowledge also the Supercomputing facility `PARAM-Shakti' of IIT Kharagpur established under National Supercomputing Mission (NSM), Government of India and supported by Centre for Development of Advanced Computing (CDAC), Pune. The sensitivity predictions related part was carried out on PARAM-Shakti.


\section*{Data Availability}
The data underlying this article will be shared on a reasonable request to the corresponding author.


\bibliographystyle{mnras} 
\bibliography{refs}




\vfill
\bsp

\label{lastpage}

\end{document}